\documentclass{article}
\usepackage{arxiv}
\usepackage[utf8]{inputenc} % allow utf-8 input
\usepackage[T1]{fontenc}    % use 8-bit T1 fonts
\usepackage{url}            % simple URL typesetting
\usepackage{booktabs}       % professional-quality tables
\usepackage{amsfonts}       % blackboard math symbols
\usepackage{nicefrac}       % compact symbols for 1/2, etc.
\usepackage{microtype}      % microtypography
\usepackage{lipsum}		% Can be removed after putting your text content
\usepackage{graphicx}
\usepackage{doi}
\usepackage[utf8]{inputenc} % allow utf-8 input
\usepackage[T1]{fontenc}    % use 8-bit T1 fonts  
\usepackage{url}            % simple URL typesetting
\usepackage{float}
\usepackage{booktabs}       % professional-quality tables
\usepackage{amsfonts}       % blackboard math symbols
\usepackage{nicefrac}       % compact symbols for 1/2, etc.
\usepackage{microtype}      % microtypography
\usepackage{lipsum}
\usepackage{fancyhdr}       % header
\usepackage{graphicx}       % graphics
\usepackage{amsmath}
\usepackage{caption}
\usepackage{hyperref}
\title{Quasi Crystal based Circular Patch Antenna with Artificial Magnetic Conductor for Breast Cancer Detection}

\author{ 
    Vishnupriya Leena \\
  Research scholar \\
  National Institute of Technology Calicut \\
  Kozhikode, Kerala\\
 \texttt{vishnupriya\textunderscore p220289ec@nitc.ac.in} \\
	\And
	 Nikhil Kumar \\
  Assistant professor \\
 National Institute of Technology Calicut \\
   Kozhikode, Kerala\\
  \texttt{nikhilkumarcs@nitc.ac.in} \\
}

\date{}

\renewcommand{\shorttitle}{\textit Quasi Crystal based Patch Antenna for Breast Cancer Detection}

\hypersetup{
pdftitle={Quasi Crystal based Circular Patch Antenna with Artificial Magnetic Conductor for Breast Cancer Detection},
pdfsubject={physics.ins-det, physics.med-ph},
pdfauthor={Vishnupriya Leena, Nikhil Kumar},
pdfkeywords={Artificial Magnetic Conductor (AMC), Penrose Pattern, Quasi crystal, Specific Absorption Rate (SAR)}
}

\begin{document}
%% Title
\title{Quasi Crystal based Circular Patch Antenna with Artificial Magnetic Conductor for Breast Cancer Detection
%%%% Cite as
%%%% Update your official citation here when published 
%\thanks{\textit{\underline{Citation}}:
\\
}

\maketitle

\begin{abstract}
Design and simulation of Quasi Crystal based Artificial Magnetic Conductor (AMC) Circular Microstrip Patch Antenna for tumor detection is discussed. Initially, a Circular Microstrip patch antenna is designed with a resonant frequency of $2.45 \,\-\text{GHz}$. The performance of the designed antenna can be improved by incorporating AMC in order to avoid surface waves. In this paper, we propose a novel technique for detecting malignant tissues in a more focusing manner using a Penrose patterned Quasi crystal. Quasi crystals act as a focusing device as it can transmit  electromagnetic waves in a narrow direction. The tumor of size $3 \, \text{mm}$ is placed at a particular location specifically at $(0,0,10)$ in the breast phantom model. Our aim is the proposed model could identify the exact tumor location or the nearest co-ordinates of the actual location. Scattering parameters $S_{11}$ and $S_{22}$ obtained at $-10.30 \, \text{dB}$ and $-9.15 \, \text{dB}$ respectively at the tumor location $(0,-4,30)$ without the presence of Quasi crystal. Results were compared in each stage with and without Penrose Quasi crystal based patch antenna. The Quasi crystal based model was simulated and the results $S_{11}$ and $S_{22}$ parameters got improved at $-10.02 \, \text{dB}$ and $-9.37 \, \text{dB}$ respectively at the tumor location $(-1, -1, 13)$. The total specific absorption rate (SAR) of the proposed model is $0.139 \, \text{W/kg}$. Hence the proposed model is highly bio-compatible and safely applicable for human body. 

\end{abstract}

% keywords can be removed
\keywords{Artificial Magnetic Conductor (AMC) \and Penrose Pattern \and Quasi crystal \and Specific Absorption Rate (SAR)}

\section{Introduction}
The demand on electronics systems for cancer diagnosis and treatment is increasing day by day. As per the survey of National Center for Health, 1,958,310 new cancer cases and 609,820 cancer deaths are projected to occur in the United States itself in 2023 \cite{siegel-2023}. Among these, Breast Cancer (Invasive Ductal Carcinoma) is the most prevalent one among women. Breast cancer can be diagnosed in certain stages \cite{ACS2022}. Stage 0 is a non invasive stage of breast cancer. It is difficult to recognize and most probably no signs of cancer cells at this stage. At Stage 1, the cancer cells are of size less than $2 \, \text{cm}$ in diameter. This is known as Invasive stage. As cancer progresses the cell properties varies such as elasticity or stiffness, thermal properties, electrical properties change accordingly. It is pointed out that the early detection and diagnosis techniques help in the survival of such cases thereby reducing the mortality rate upto a certain extend. The primary key in women’s survival from breast cancer is early detection and proper treatment \cite{abdul2021existing}. Various breast cancer imaging techniques were extracted as follows: Mammography, Contrast-enhanced Mammography, Digital Tomosynthesis, Sonography, Sonoelastography, Magne\-tic resonance imaging (MRI), Magnetic Elastography,  Diff\-usion- weighted imaging, Magnetic spectroscopy, Ultrasound, Nuclear medicine, Optical imaging, and Microwave imaging \cite{iranmakani2020review,bhushan2021current}.

Mammography is the most commonly used diagnosing technique for early diagnosis of breast cancer. The most commonly cited risks of mammography screening are over-diagnosis, false-positives, anxiety, and radiation injury \cite{jahan2021microstrip}. The ionizing nature of X-rays and the discomfort felt by women due to the pressure exerted on their breasts for imaging, discourage them from regular checkups \cite{abu2014designing,simovski2005high}. In Ultrasound imaging, the quality of the images is so low that they cannot clearly distinguish between a normal cell and a malignant cell in its early stage \cite{simovski2005high,hosseini2006novel}. MRI is a more sensitive technique that can be used for women with dense breasts because of its high sensitivity \cite{hamza2022low}. However, it is very expensive. In addition, the breast cannot be positioned correctly in this technique, which may lead to an incorrect diagnosis \cite{abu2014designing,dewan2017artificial}. Due to the frequent inaccuracies and testing constraints associated with the previously described methods, researchers are now considering a novel microwave-based technology. Most of the detection techniques discussed here now in use are of invasive and painful. Patients with metal implants, pacemakers are restricted to undergo such investigations. Also, these techniques are quite expensive for common man \cite{sabila2022design}. Microwave imaging technique (MWI) is the best alternative compared to earlier techniques. The advantages of MWI are its higher data rate accuracy, low cost, non-ionizing nature, reduced complexity, comfortable positioning, and very low power density \cite{sievenpiper1999high,hamza2022low}. 

MWI is a new trend and achieved a remarkable attraction in the field of cancer diagnosis. The basic principle of the MWI technique is to analyze and distinguish between changes in the back scattered signal and in the different electrical properties of cells and tissues \cite{dewan2017artificial,bhargava2022microwave,hamza2022low}.
Antennas play a major role in the MWI \cite{balanis2016antenna}. It can transmits microwave signal and measures the back scattered signal \cite{wang2023microwave}. Microstrip patch antennas or printed antennas (MPAs) are now become a common trend in the field of Industrial, Scientific and Medical applications (ISM). Advantages of Microstrip antennas are low fabrication cost, light weight, low volume, and low profile configuration that can be made conformal, it can be easily mounted on rockets, missiles and any conformal shaped satellites without major modifications and arrays of these antennas can simply be produced \cite{el2020survey} . In this paper, a Circular Microstrip patch antenna resonating at a frequency of $2.45 \, \text{GHz}$ is used. However, there are certain limitations for the patch antenna like smaller bandwidth, higher cross polarization levels and lower gain. These limitations can be improved using different techniques such as Photonic Band gap (PBG), Defected Ground Structure (DGS), Electromagnetic Band gap (EBG) etc. \cite{sabila2022design,jahan2021microstrip,el2020survey}. Defected ground structure (DGS) is commonly used to improve the performance of traditional MPAs between $10 \, \text{dB}$ bandwidth and gain \cite{balanis2016antenna,areed2023breast}. This feature has been used in the proposed design to improve the various patch antenna's performances such as wide bandwidth, high gain, and efficiency \cite{areed2023breast}. The DGS on the ground plane increases the fringing field which introduces parasitic capacitance. This parasitic capacitance increases coupling between  the conducting patch and the ground plane which is responsible for the enhancement of the bandwidth \cite{sabila2022design,7095318}.

From the recent studies, Metamaterials have been used as an effective medium to enhance the conventional antenna’s gain and directivity \cite{dewan2017artificial,abu2014designing,sievenpiper1999high}. Artificial Magnetic Conductors (AMC) are such type of metamaterial-based structures that can reduce back scattering and detect the cancerous cells by focusing in a particular direction. AMC is a high impedance surface material, hence its reflection phase is zero. Circular patch antenna combine with AMC shown better improvement in the directivity. Most of the diagnosing techniques are often dangerous to surrounding tissues especially hyperthermia, chemotherapy and  radiotherapy \cite{ACS-chemotherapy,cancernet-longterm}. Thus, a novel design of antenna which is highly focusing for cancer cells and less harm to neighbouring cells is also now in demand in the medical fields. From the literature review done, Quasi crystals meet the requirement. An aperiodic pattern named Penrose style tiling is used for designing the Quasi crystal. In this paper, Circular patch antenna incorporated with AMC and Penrose Quasi crystal is discussed.

This work proposes a novel design of Circular patch antenna with AMC and Penrose style Quasi crystal for cancer diagnosis and treatment. Photonic crystals and Metamaterials can manipulate light in ways that cannot be achieved with uniform materials \cite{neve-oz2010resonant}. However, Photonic Quasi crystals can possess large omnidirectional band gaps because of their higher rotational symmetry \cite{ieee1992ieee}. Instead of using Quasi crystals lens, a two-dimensional pattern of metamaterial was arranged in the form of kites and darts \cite{bhargava2022microwave}. It acts as a transmitting medium of electromagnetic wave from the antenna to the receiver section. Specific Absorption Rate (SAR) and reflection coefficient analyzed for locating the tumor. SAR measures electromagnetic radiation absorption by tissues and represents the amount of energy per unit mass of biological tissue. Aim of this paper is to i) detect the tumor and identify its location, ii) calculate the Specific Absorption Rate (SAR). The results demonstrate the performance of various configurations of microstrip patch antennas integrated with AMC and Quasi crystals, using a breast phantom model for both normal (healthy) breast and breast with tumor.

This paper is organized as follows: Section 2 explains about the design and analysis of Circular patch antenna and AMC. Section 3 describes about the proposed model. Simulation results and performance comparisons are discussed in the Section 4. Section 5 refers to the conclusion and future scope of this work.

\section{Design of Circular Microstrip Patch Antenna and AMC}
\label{sec:1}
Circular patch antenna consists of a ground plane, substrate layer and a circular patch at the top of the structure shown in Fig. 1. Ground plane and circular patch is made up of copper because copper is an excellent conductor and is highly reactive. It can distribute electrical energy effectively. Besides copper is much harder and relatively cheap. FR-4 lossy is chosen as the substrate material due to its availability and low cost \cite{el2020survey}. Circular patch antenna is designed at $2.45 \, \text{GHz}$ using equation \cite{sabila2022design}.

\begin{equation}
	a= \frac{F}{\{{1+\frac{2h}{\pi\epsilon r}[ln(\frac{\pi F}{2h})+1.7726]\}^{0.5}}}
\end{equation}

The value of $F$ for (1) can be calculated using equation (2)

\begin{equation}
	F=\frac{8.791 \cdot 10^9}{f_r\sqrt{\epsilon_r}}
\end{equation}

\begin{figure}[htbp]
	\centering
	\includegraphics[width=0.5\linewidth]{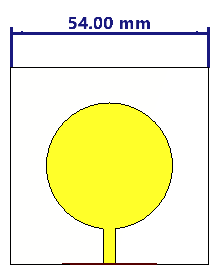}
	\caption[]{Circular patch antenna}
	\label{fig:1}
\end{figure}

 At first a single circular patch antenna with a port 1 is simulated in the free space. Reflection coefficient or $S_{11}$ represents how much power an antenna can reflect. Obtained scattering parameter ($S_{11}$) and radiation pattern of Circular patch antenna is shown in Fig. 2 and Fig. 3 respectively. Results shown that S-parameter ($S_{11}$) of circular patch antenna was obtained below $-11.6 \, \text{dB}$. 

\begin{figure}[H]
	\centering
	\includegraphics[width=0.5\linewidth]{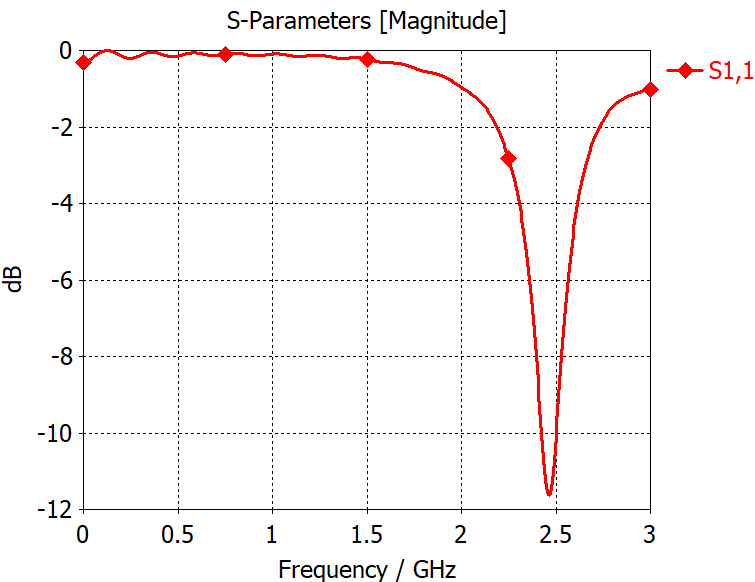}
	\caption{Simulated reflection coefficient $S_{11}$ }
	\label{fig:2}
\end{figure}
\begin{figure}[H]
	\centering
	\includegraphics[width=0.5\linewidth]{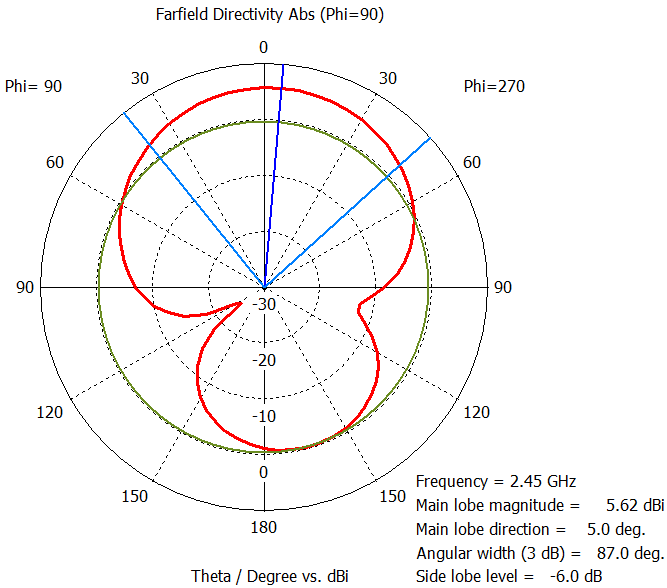}
	\caption{Simulated radiation pattern 1D plot}
	\label{fig:3}
\end{figure}

% Antenna Parameter Table
\begin{table}[H]
\centering
	\caption{Antenna Parameters and Dimension}
	\label{tab:tbl1}      
	\begin{tabular}{lll}
		\hline\noalign{\smallskip}
		Parameter & Dimension(mm)  \\
		\noalign{\smallskip}\hline\noalign{\smallskip}
			Ground Length (Lg) & 54  \\
		Ground Width (Wg) & 54 \\
		Substrate Length (Ls) & 54  \\
		Substrate Width (Ws) & 54  \\
		Substrate Height (h) & 1.5 \\
		Patch radius (a) & 17.28 \\
		Feed Length (lf) & 3.12 \\
		\noalign{\smallskip}\hline
	\end{tabular}
\end{table}

\subsection{Significance of AMC}

Radiation pattern obtained for a Circular patch antenna is broader and has back radiation as depicted in Fig. 3. The main reason is the effect of surface waves. This will adversely affect the tumor detection. Surface waves can be reduced by using Artificial Magnetic Conductor (AMC). Side lobe level of AMC placed antenna is less compared to the conventional antenna. AMC is a metamaterial which replicates the characteristics of a perfect magnetic conductor (PMC). Metamaterials are artificial structures that are realized by embedding metallic material with periodic pattern onto the dielectric substrate \cite{abu2014designing,sievenpiper1999high,simovski2005high,hosseini2006novel}. The ability of a PMC to provide zero-degree reflection phases at its resonance frequency is a useful trait that AMC also imitates. AMC is a high impedance surface material, hence its reflection phase is zero. There is less wasted power in the back lobe, and the radiation pattern is smoother. Because of its in-phase characteristics, the AMC will cancel out the source's current with its image \cite{balanis2018applications}. Generally, an AMC is referred to as a high impedance surface (HIS) when it is utilized as an antenna ground plane. Another peculiar property of AMC is that its tangential magnetic field is zero at the surface. A detailed study of gain enhancement of AMC based millimeter wave microstrip patch antenna is discussed in \cite{belabbas2021artificial}.

\begin{figure}[H]
	\centering
	\includegraphics[width=0.5\linewidth]{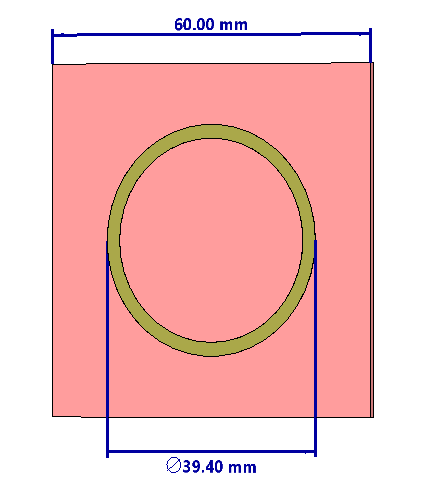}
	\caption{Unit cell AMC}
	\label{fig:antenna-and-amc}
\end{figure}

\subsection{Design of AMC}
The designed AMC layout is shown in the Fig. 4. The unit cell of AMC is designed by using the concept of group 2 FSS circular loop structure whose outer loop having radial length of $\lambda$/3 \cite{anwar2018frequency}. The perspective view of designed AMC unit cell is shown in Fig. 5. 

\begin{figure}[H]
	\centering
	\includegraphics[width=0.5\linewidth]{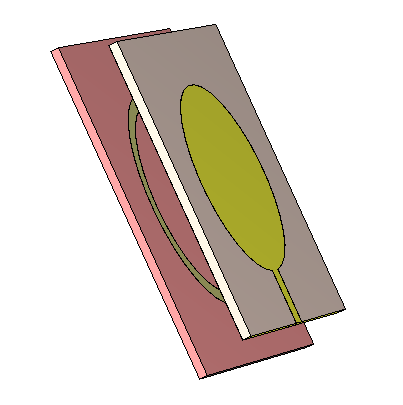}
	\caption{Antenna with AMC}
	\label{fig:antennaamc}
\end{figure}

\begin{figure}[H]
	\centering
	\includegraphics[width=0.5\linewidth]{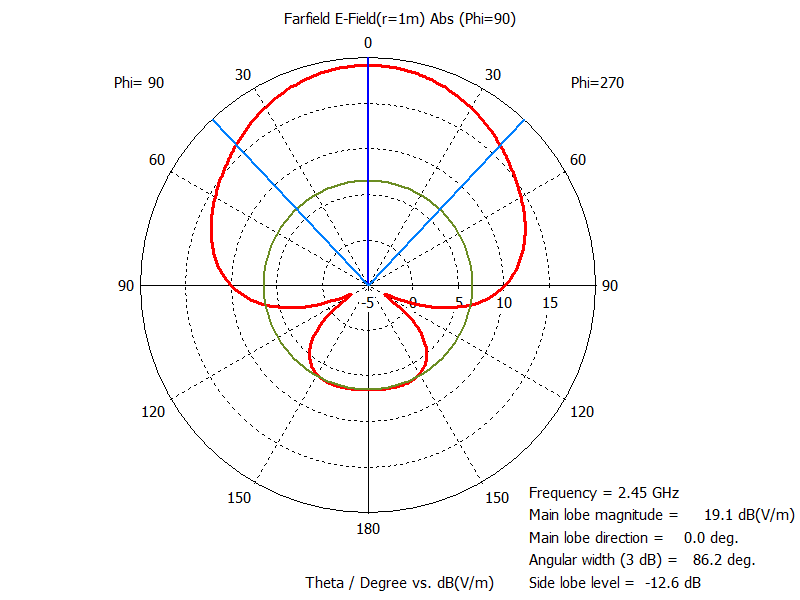}
	\caption{Simulated radiation pattern of AMC antenna}
	\label{fig:5}
\end{figure}

AMC consists of a substrate layer at the bottom and at the top is Frequency selective surface (FSS). FR-4 lossy with dielectric constant $\epsilon = 4.3$, a dielectric loss tangent $\tan \delta$ of $0.025$ and thickness of $1.5 \, \text{mm}$ is used as the substrate material. FSS is a periodic surface with identical two-dimensional arrays of elements arranged on a dielectric substrate. AMC's are based on periodically engineered structures that act as a sheet of magnetic currents having magnetic polarizability \cite{anwar2018frequency}. AMC is placed below the antenna which acts as a ground. Simulated radiation pattern is shown in Fig. 6. Compared to Fig. 3, it is clear that AMC based antenna improves the radiation in a particular direction and reduces the broadening of the pattern. AMC is an implicit method for improving antenna performance in terms of gain, directivity, bandwidth, and radiation. It is a type of artificially created periodically loaded structure that would be used to limit antenna propagation in a particular direction or reduce backward radiation, also known as phase reflection \cite{zerrad2022multilayered}. This phase reflection is the feature caused by an AMC that is not found in nature. Hence, this method is truly applicable for the detection and location of cancer cells. 

\section{Proposed model}
The proposed model comprises a circular patch antenna with AMC and a Penrose style Quasi crystal covering the breast phantom model. AMC design and features are already discussed in the previous section. 

\begin{table}[H]
\centering
\caption{Electrical Properties of breast phantom model}
\label{tab:tabl2}     

\begin{tabular}{llll}
\hline\noalign{\smallskip}
Tissue & Permittivity  & Conductivity & Density  \\
\noalign{\smallskip}\hline\noalign{\smallskip}
Skin & 36.7 & 2.34 & 1109 \\
Adipose Tissue  & 4.84 & 0.262 & 911 \\
Glandular Tissue & 50 & 3.46 & 1041 \\
Tumor & 54.9 & 4 & 1058 \\
\noalign{\smallskip}\hline
\end{tabular}
\end{table}

\begin{figure}[H]
	\centering
	\includegraphics[width=0.5\linewidth]{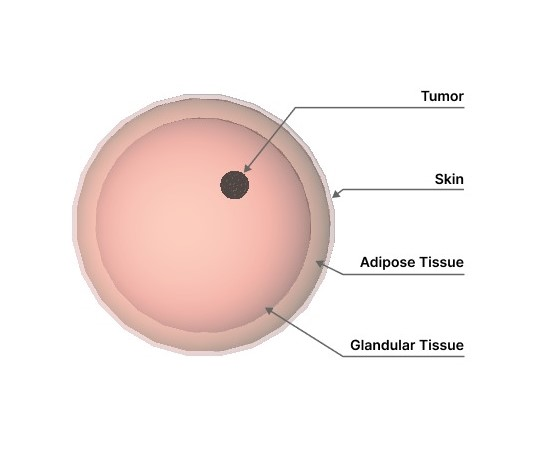}
	\caption{Breast phantom model}
	\label{fig:6}
\end{figure}

\begin{figure}[H]
	\centering
	\includegraphics[width=0.5\linewidth]{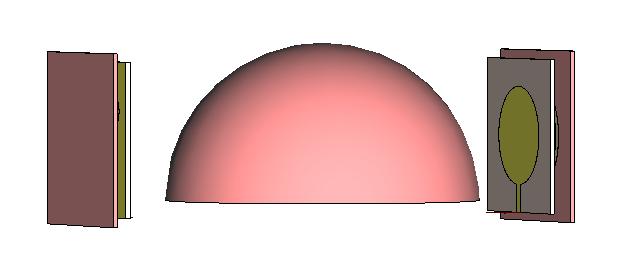}
	\caption{Proposed model without QC}
	\label{fig:7}
\end{figure}

The breast phantom model is designed using CST Studio Suite with the electrical properties mentioned in the Table 2 \cite{9733569}. The designed breast phantom model is shown in the Fig. 7. The designed model, Circular patch antenna with AMC, and breast phantom model represented in Fig. 8.

\section{Simulation results and Performance comparisons}
At first the breast phantom model is placed between antenna which acts a transmitter and receiver. The antenna is radiated towards the breast phantom model for locating the tumor cell. The location of the tumor is identified by analyzing the reflection coefficient or S-parameters and rate of absorption is calculated by another parameter known as Specific Absorption Rate (SAR). All simulations were performed using CST Studio Suite. Reflection coefficients $S_{11}$ and $S_{22}$ for the model without tumor was simulated. From Fig. 9, For $S_{11}$, the value of reflection coefficient is obtained at $-10.14 \, \text{dB}$ and for  $S_{22}$, $-9.00 \, \text{dB}$. The dielectric difference found in the healthy and cancerous tissues not only produce difference in the reflection coefficients but also shows difference in the way they absorb those incident microwave radiation \cite{bhargava2022microwave}. Fig. 10 shows the results obtained with the presence of tumor in the breast phantom model for the corresponding S-parameters $S_{11}$ and $S_{22}$  obtained at $-10.30 \, \text{dB}$ and $-9.30 \, \text{dB}$ respectively. The results showed that the proposed antenna could locate the presence of tumor as the reflection coefficients are different for both conditions. S-parameters value for tumor cells is lower compared to normal tissue.

\begin{figure}[H]
	\centering
	\includegraphics[width=0.5\linewidth]{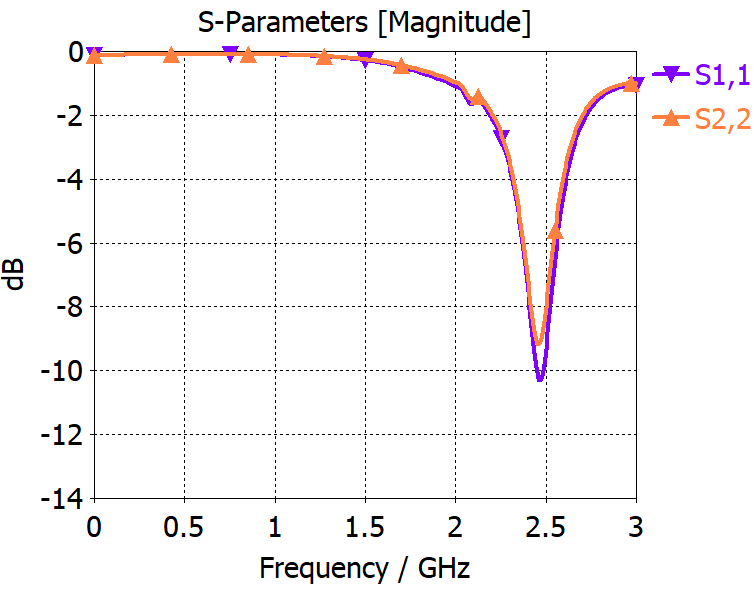}
	\caption{S-parameters obtained for normal tissue without Quasi crystal}
	\label{fig:8}
\end{figure}
\begin{figure}[H]
	\centering
	\includegraphics[width=0.5\linewidth]{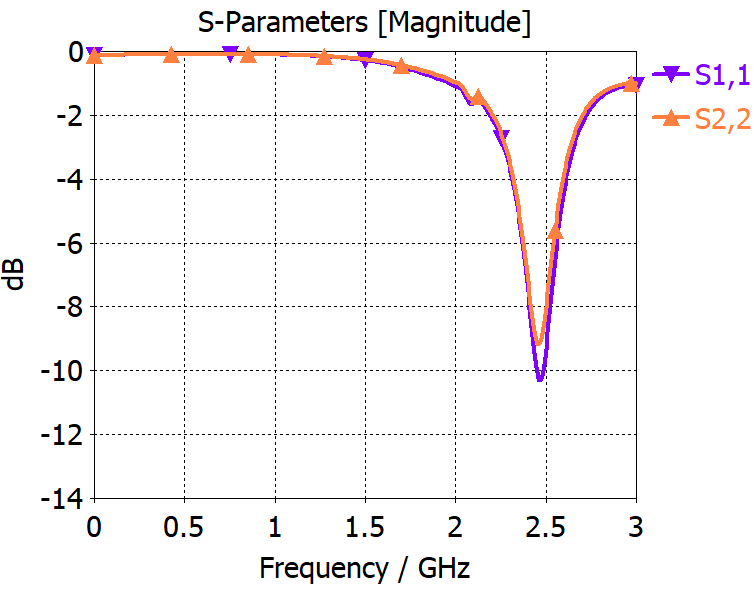}
	\caption{S-parameters obtained for tumor without Quasi crystal}
	\label{fig:9}
\end{figure}

From the literature review, there are lot of researches done for the breast cancer detection. Many papers pointed out only the presence of tumor. The main focus of this work is to detect the actual location of the tumor. The physics behind the periodic system is fundamentally important because they give rise to a number of phenomena that control wave transmission and interference. Deviations from periodicity can lead to increased complexity and a variety of unexpected outcomes. In the science of optics, one example of a deviation from the norm is the development of Photonic Quasi crystals, a class of structures composed of building blocks placed according to carefully thought-out designs, but lacking translational symmetry. Nevertheless, these structures, which lie between periodic and disordered structures, still show sharp diffraction patterns that confirm the existence of wave interference resulting from their long-range order \cite{vardeny2013optics}. The major advantages of Quasi crystals are i) can enhance wave transmission and boost non linear phenomena due to rotational symmetry forbidden in  conventional periodic crystals ii) can absorb electromagnetic wave compared with corresponding periodic absorbers \cite{ma2022quasiperiodic}. Quasi crystals offer a superior flexibility over regular magnonic crystals for band structure engineering \cite{lisiecki2019magnons}. A Penrose styled Quasi crystal is introduced here. Non-periodic tiling, like Penrose tiling, are important models for physical Quasi crystals. A model of Penrose tiling generated by MATLAB functions is shown in the Fig. 11. It is made up of an anisotropic Quasi metasurface with dielectric constant 2.2. 

\begin{figure}[H]
	\centering
	\includegraphics[width=0.5\linewidth]{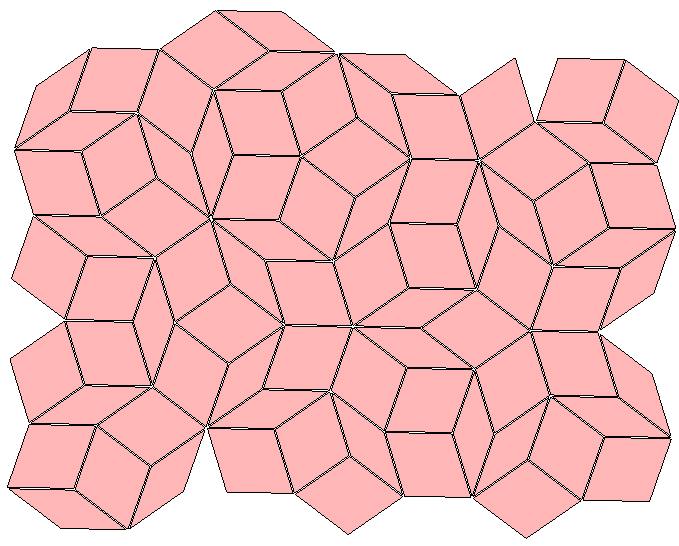}
	\caption{2D Penrose Quasi crystal with kites and darts }
	\label{fig:10}
\end{figure}

The structure of a Quasi crystal is defined as follows. A structure $\rho(r)$ is a Quasi crystal if and only if its diffraction intensity function \cite{edagawa2014photonic},

\begin{equation}
	I(S) =  \lvert F(S) \rvert ^2 = \lvert \frac{1}{V} \int{\rho(r)} \exp(-i S.r) \rvert^2
	\end {equation}
	satisfies the following two conditions: (i) indicates that Photonic Quasi crystals exhibit Bragg scattering of light (ii) indicates that Photonic Quasi crystals can possess a higher rotational symmetry than conventional Photonic crystals.
	Photonic crystals with a complete PBG can act as a mirror that selectively reflects light of a particular frequency (or wavelength) range. At the same time, they can also act as a filter that transmits light of a particular frequency (or wavelength) range outside the PBG \cite{edagawa2014photonic}. Aperiodic tiling, like Penrose tiling, are important models for physical Quasi crystals. A tile-substitution is given by a set of prototiles $T_{1}$, . . . , $T_{m}$, a substitution factor $\lambda$, and a rule how to replace the enlarged prototiles $\lambda \cdot T$ with congruent copies of the prototiles \cite{oh2016fractal}. The matching rules for kites and darts involve geometric relationships that can be expressed using Fibonacci sequences. The Penrose tiling exhibits a five-fold symmetry, and can be seen as a two-dimensional model of a quasi-crystal \cite{oh2016fractal,neve-oz2010resonant,sbordone-2023}. It can be modelled as an aperiodic “disjoint” covering of the plane generated by two rhombi $R_{36}$$^{\circ}$ and $R_{72}$$^{\circ}$ with equal lengths \cite{jaric2012introduction}.
	
	\begin{equation}
	\phi = \frac{\text{area } R_{72^\circ}}{\text{area } R_{36^\circ}} = \frac{1+\sqrt{5}}{2}\text{(golden ratio)}\quad(\phi^2-\phi-1 = 0)
	\end{equation}
	
	By placing the quasi crystal in front of the breast phantom it can enhance the wave transmission and thereby locate the tumor compared to existing models. The narrow beam focusing using quasi crystals is associated with negative refraction \cite{ieee1992ieee} . A study of the AMC based antenna and Quasi crystal included in order to point out the significance of Quasi crystal for the improvement of S-parameters is included. For a single port antenna, S-parameter $S_{11}$ obtained at $-13.8 \, \text{dB}$ which is a better result compared to Fig. 2. 
	
\begin{figure}[H]
	\centering
	\includegraphics[width=0.5\linewidth]{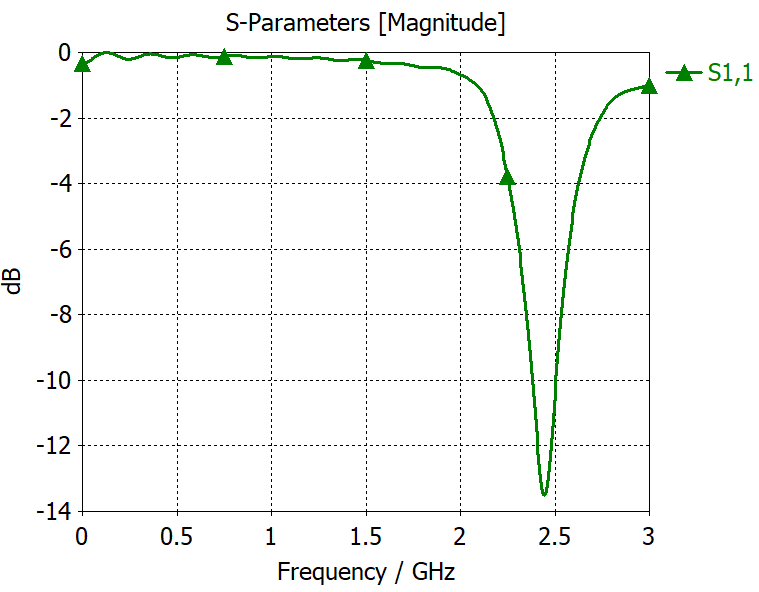}
	\caption{S-parameters $S_{11}$ and $S_{22}$ obtained with Quasi crystal}
	\label{fig:11}
\end{figure}

\begin{figure}[H]
		\centering
		\includegraphics[width=0.5\linewidth]{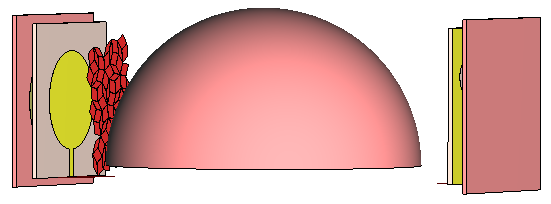}
		\caption{Proposed model with Quasi crystal}
		\label{fig:12}
	\end{figure}

\subsection{Analysis of Scattering parameters and SAR}

	Fig. 13 depicts the proposed model. A Penrose Quasi crystal acts as the medium of transmission between the antenna with AMC and the breast phantom model. Quasi crystal is arranged in the pattern of Penrose tiles as shown in Fig. 11. Scattering parameters were compared with the new model and existing model. Results show that the Scattering parameters $S_{11}$ and $S_{22}$ obtained with the presence of Quasi crystal is $-10.2 \, \text{dB}$  and $-9.37 \, \text{dB}$ shown in Fig. 12. When comparing with the model without Quasi crystal depicted in Fig. 10, $S_{11}$ and $S_{22}$ values are $-10.29 \, \text{dB}$ and $-9.15 \, \text{dB}$ respectively. The model could locate the tumor at $(-1,-1,13)$ which is closer to the actual tumor location $(0,0,10)$. This result shows a significant improvement as the S-parameters as well as tumor location in the presence of Quasi crystal.
	
	Now the next step is to calculate the absorption rate of EM radiation by tissues in Specific Absorption rate \cite{bhargava2022microwave}. The term Specific Absorption Rate describes the amount of energy which is absorbed (W/kg) in the breast tissue \cite{ieee1992ieee}. It can be expressed as,
	
	\begin{equation}
		SAR = \frac {\sigma |E|^2} {\rho}, 
	\end{equation}
	Where $\sigma$ is the conductivity of the material in S/m, E is the electric field intensity in $V/m$ and $\rho$ is the mass density in $Kg/m^3$ \cite{9733569}.
	
	Usually SAR is analyzed for $1g$ tissue and $10g$ tissue in the CST software. It can be represented pictorially as illustrated in Fig. 14 and Fig. 15.
	
	\begin{figure}[H]
		\centering
		\includegraphics[width=0.5\linewidth]{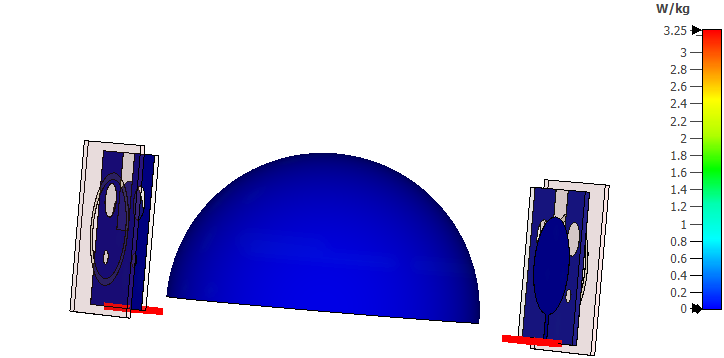}
		\caption{SAR (f=2.45) [1] ($10g$) without Quasi crystal}
		\label{fig:13}
	\end{figure}
	\begin{figure}[H]
		\centering
\includegraphics[width=0.5\linewidth]{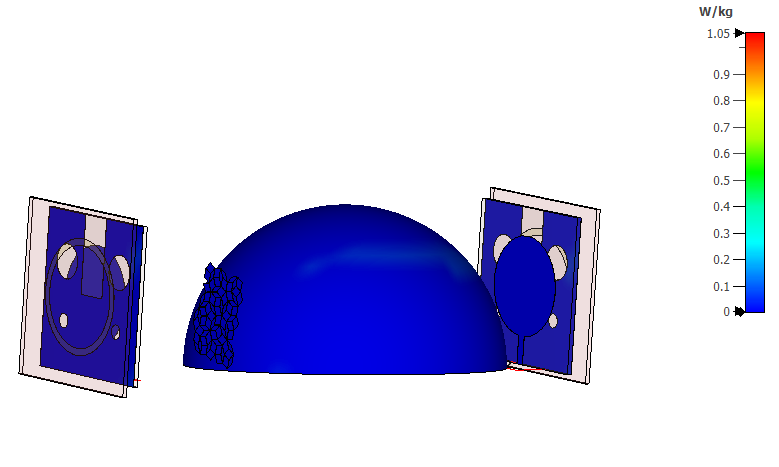}
		\caption{SAR (f=2.45) [1] ($10g$) with Quasi crystal}
		\label{fig:14}
	\end{figure}
	
	A comparison chart including all the simulation results discussed earlier is rendered here. Simulations carried out for the scattering parameters and SAR value. Importance of the Quasi crystal in this model is how much the model can locate and identify the tumor location. Tumor location is also discussed along with S-parameters and SAR value. Table 3 represents the results obtained by the designed model without quasi crystal. $S_{11}$ and $S_{22}$ are $-10.9 \, \text{dB}$ and $-9.15 \\ \text{dB}$ obtained respectively for this case. Without placing Quasi crystal, the tumor was identified at $(0,-4,30)$ which is far away from the actual location. The proposed model with Quasi crystal identified the tumor at $(-1,-1,13)$ which is near to the exact location. SAR value obtained for $10g$ tissue is higher for model without Quasi crystal which was simulated at $3.25 \, \text{W/kg}$ and $6.09 \, \text{W/kg}$ for $10g$ and $1g$ tissue respectively. While SAR rate reduced to $1.05 \, \text{W/kg}$ for the proposed model which is illustrated in the Table 4. Total SAR calculated for both models. From the comparison chart , it is clear that total SAR value for Quasi crystal based model is lower compared to existing model. When comparing both results, SAR rate for $10g$ tissue and total SAR is lower for the new proposed model. Since the rate of EM absorption is less , this model is bio compatible. Bio-compatible defines this device will cause less harm to the normal human tissues \cite{ieee1992ieee}. 
	\begin{table}[H]
    \centering
    \caption{Results obtained with Tumor and without Quasi Crystal}
    \label{tab:2}     
    \begin{tabular}{|p{1.5cm}|p{1.5cm}|p{2.5cm}|p{1.5cm}|p{2.5cm}|}  
        \hline
        $\mathbf{S}_{11}$ (dB) & $\mathbf{S}_{22}$ (dB) & 
        \textbf{SAR ($10g$)} & \textbf{SAR ($1g$)} & \textbf{Tumor Location}  \\
        \hline
        -10.29 & -9.15  & 3.25 & 6.09 & $(0,-4,30)$ \\
        \hline
    \end{tabular}
    
    \vspace{5pt}  % Adds spacing before extra text
    {\small Tumor size = $3 \, \text{mm}$ and location $(0,0,10)$.} % Smaller font for additional info
\end{table}

\begin{table}[H]
    \centering
    \caption{Results Obtained with Tumor and Quasi Crystal}
    \label{tab:3}
    \begin{tabular}{|p{1.5cm}|p{1.5cm}|p{2.5cm}|p{1.5cm}|p{2.5cm}|}  
        \hline
        $\mathbf{S}_{11}$ (dB) & $\mathbf{S}_{22}$ (dB) & \textbf{SAR ($10g$)} & \textbf{SAR ($1g$)} & \textbf{Tumor Location}  \\
        \hline
        -10.02 & -9.37  & 1.05 & 2.4 & $(-1,-1,13)$ \\
        \hline
    \end{tabular}
    
    \vspace{5pt}  % Adjust spacing before additional information
    {\small Tumor size = $3 \, \text{mm}$ and location $(0,0,10)$.}
\end{table}

\section{Conclusion}
A Quasi crystal based Circular microstrip patch antenna with Artificial Magnetic Conductor is designed and simulated at a frequency of 2.45 GHz. Simulation results include scattering parameters, directivity, Specific Absorption Rate (SAR). Simulations were carried out for detecting the tumor location with and without the presence of Quasi crystal. A 2D model of Penrose Quasi crystal is placed in between the transmitting antenna and the breast phantom model. A tumor size of $3 \, \text{mm}$ was used within the breast phantom model. The proposed model could locate the tumor at $(-1,-1,13)$ which is closer to the actual tumor location $(0,0,10)$. From the results obtained, it is clear that the parameters varies accordingly with and without the presence of Quasi crystal and thus the proposed model is highly recommended as a novel microwave imaging technique. This model could locate the tumor nearer to the maximum coordinates of the actual location where the tumor placed. SAR value obtained for the tumor or malignant cell is higher compared to normal breast tissue. Lower SAR value for the proposed model proves that the model is safely applicable for human tissues. In future, this model can be extended for high frequencies and can be applicable for hyperthermia treatment also. 

\subsection*{\textbf{Author Contributions}}
 N.K. conceived the conceptualization and formalization of research and coordinated the project. V.L performed modeling and simulation of a Quasi crystal based antenna system for cancer detection. V.L wrote the initial draft. V.L and N.K analyzed the data and reviewed the manuscript.
\section*{Declarations}

\noindent \textbf{Conflict of Interest:} The authors have no relevant financial or non-financial interests to disclose.

\vspace{6pt} 

\noindent \textbf{Ethical Approval:} Not applicable.
\bibliographystyle{unsrt}
\bibliography{references}

\end{document}